\documentstyle[12pt]{article}
\newcommand{\beq}{\begin{equation}}
\newcommand{\eeq}{\end{equation}}
\newcommand{\beqa}{\begin{eqnarray}}
\newcommand{\eeqa}{\end{eqnarray}}
\newcommand{\ba}{\begin{array}}
\newcommand{\ea}{\end{array}}

\begin{document}

\begin{flushright}
To be published in \\
Open Systems and Information Dynamics
\end{flushright}
\vskip 0.5 truecm

\begin{center}
{\large \bf Semiclassical Expansion for the Angular Momentum}
\footnote{Presented as a short report at the 3rd International 
Summer School/Conference {\it Let's Face Chaos Through Nonlinear Dynamics}, 
24 June -- 5 July, 1996, Maribor, Slovenia.} \\
\vspace{0.3in}
MARKO ROBNIK$^{(*)}$ and LUCA SALASNICH$^{(*)(+)}$ \\
\vspace{0.2in}
$^{(*)}$ Center for Applied Mathematics and Theoretical Physics,\\
University of Maribor, Krekova 2, SLO--2000 Maribor, Slovenia\\
\vspace{0.2in}
$^{(+)}$ Dipartimento di Matematica Pura ed Applicata \\
Universit\`a di Padova, Via Belzoni 7, I--35131 Padova, Italy \\
\end{center}

\vspace{0.3in}

\normalsize 
After reviewing the WKB series for the Schr\"odinger equation 
we calculate the semiclassical expansion for the 
eigenvalues of the angular momentum operator. 
This is the first systematic semiclassical treatment of the angular 
momentum for terms beyond the leading torus approximation. 

\section{Introduction}

The semiclassical method of torus 
quantization is just the first term of a certain $\hbar$-expansion, 
usually called the WKB expansion. 
The method goes back to the early days of quantum mechanics and 
was developed by Bohr and Sommerfeld for one-freedom systems and
separable systems, it was then generalized for integrable 
(but not necessarily separable) systems by Einstein [1], 
which is called EBK or torus quantization. In fact, Einstein's
result was corrected for the phase changes due to caustics by Maslov [2,3], 
but the torus quantization formula thus obtained is still
just a first term in the WKB expansion, whose higher terms 
can be calculated with a recursion formula in one degree systems, 
but are generally unknown in systems with more than one degree of freedom. 
\par
Our goal in the present paper is to generalize the WKB expansion 
of the Schr\"odinger equation to the angular momentum operator. 
To the best of our knowledge a detailed analysis
of this problem has not been undertaken in the literature
so far. Thus our present work is the first systematic semiclassical
expansion of the angular momentum problem. 
\par
In section 2 we treat the one--dimensional stationary Schr\"odinger 
equation by analyzing the corrections to the leading 
torus quantization term, and in 
the section 3 we study the solutions of the angular momentum 
operator by calculating the corrections to the
leading torus quantization term. In section 4 we discuss
the results and draw some general conclusions.

\section{WKB expansion for the Schr\"odinger equation}

We consider the one--dimensional stationary Schr\"odinger equation 
\beq
\big( -{\hbar^2 \over 2} {d^2 \over dx^2} + V(x) \big) \psi (x) = E \psi (x) 
\; ,
\eeq
where $V(x)$ is a potential with two turning points. 
We can always write the wave function as
\beq
\psi (x) = \exp{ \big( {i\over \hbar} \sigma (x) \big) } \; ,
\eeq
where the phase $\sigma (x)$ is a complex function that satisfies 
the differential equation
\beq
\sigma{'}^2(x) + ({\hbar \over i}) \sigma{''}(x) = 2 (E - V(x)) \; .
\eeq
The WKB expansion for the phase is
\beq
\sigma (x) = \sum_{k=0}^{\infty} ({\hbar \over i})^k \sigma_k(x) \; .
\eeq
Substituting (4) into (3) and comparing like powers of $\hbar$ gives 
the recursion relation ($n>0$)
\beq
\sigma{'}_0^2=2(E-V(x)), \;\;\;\; 
\sum_{k=0}^{n} \sigma{'}_k\sigma{'}_{n-k}
+ \sigma{''}_{n-1}= 0 \; .
\eeq
\par
The quantization condition is obtained by requiring 
the uniqueness of the wave function
\beq
\oint d\sigma = 
\sum_{k=0}^{\infty} ({\hbar \over i})^{k} \oint d\sigma_{k}=
2 \pi n \hbar \; ,
\eeq
where $n \geq 0$, an integer number, is the radial quantum number. 
\par
The zero order term, which gives the Bohr-Sommerfeld formula, 
is given by
\beq
\oint d\sigma_0 = 2 \int dx \sqrt{2 (E - V(x))} \; ,
\eeq
and the first odd term in the series gives the Maslov corrections
(Maslov index is equal to 2) 
\beq
({\hbar \over i}) \oint d\sigma_{1} = - \pi \hbar \; . 
\eeq
All the other odd terms vanish when integrated along the closed 
contour because they are exact differentials [4]. 
So the quantization condition can be written
\beq
\sum_{k=0}^{\infty} ({\hbar \over i})^{2k} \oint d\sigma_{2k} = 2 \pi 
(n +{1\over 2}) \hbar \; ,
\eeq 
thus a sum over even--numbered terms only. 
The next two non--zero terms are [4,5] 
\beq
({\hbar \over i})^{2} \oint d\sigma_2 
= - \hbar^2 {1\over 12} {\partial^2 \over \partial E^2} 
\int dx {V{'}^2(x) \over \sqrt{2 (E - V(x))} } \; , 
\eeq
\beq
({\hbar \over i})^{4} \oint d\sigma_4 = \hbar^4
\big( {1\over 240} {\partial^3\over \partial E^3} 
\int dx {V{''}^2(x) \over \sqrt{2 (E - V(x))} } 
- {1\over 576} {\partial^4\over \partial E^4} 
\int dx {V{'}^2(x) V{''}(x) \over \sqrt{2 (E - V(x))} } \big) \; .
\eeq
So we have obtained the first two quantum corrections to the torus 
quantization for the one--dimensional stationary Schr\"odinger equation. 
We note that higher--order corrections quickly increase in complexity 
and only in a few cases a systematic WKB
expansion can be worked out even explicitly to all orders, resulting in
a convergent series whose sum is identical to the exact spectrum [4,6,7]. 

\section{Semiclassical expansion for the angular momentum}

We consider the eigenvalue equation of the angular momentum
\beq
{\hat L}^2 Y(\theta ,\phi ) = \lambda^2 \hbar^2 Y(\theta ,\phi ) \; ,
\eeq
where ${\hat L}^2$ is formally given by the equation 
\beq
{\hat L}^2 = {\hat P}_{\theta}^2 + 
{{\hat P}_{\phi}^2 \over \sin^2{(\theta )}} \; , 
\;\;\;\;\;\; P_{\phi}=L_z \; ,
\eeq
with 
\beq
{\hat P}_{\theta}^2= -\hbar^2 ({\partial^2 \over \partial \theta^2} 
+ \cot{(\theta )} {\partial \over \partial \theta}) \; ,
\eeq
\beq
{\hat P}_{\phi}^2= -\hbar^2 {\partial^2 \over \partial \phi^2} \; .
\eeq
We can write the eigenfunction as
\beq
Y(\theta ,\phi ) = T(\theta ) e^{i n_{\phi} \phi} \; , 
\eeq
and we obtain
\beq
{\hat P}_{\phi}^2 Y(\theta ,\phi ) = n_{\phi}^2 \hbar^2 Y(\theta ,\phi ) 
\; ,
\eeq
and also
\beq
T{''}(\theta ) + \cot{(\theta )} T{'}(\theta ) 
+ (\lambda^2 - {n_{\phi}^2 \over \sin^2{(\theta )} })T(\theta )=0 \; .
\eeq
Notice that $\hbar$ does not appear in this equation anymore. 
To perform the WKB expansion we introduce a small parameter $\epsilon$,
which might be thought of as proportional to $\hbar$, 
and consider the eigenvalue problem
\beq
\epsilon^2 T{''}(\theta ) + \epsilon^2 \cot{(\theta )} T{'}(\theta ) 
= Q(\theta) T(\theta ) \; ,
\eeq
where 
\beq
Q(\theta )= W(\theta ) - \lambda^2 
={n_{\phi}^2 \over \sin^2{(\theta )}}-\lambda^2 \; .
\eeq
Thus small $\epsilon$ limit is equivalent to the large $n_{\phi}$ and/or 
large $\lambda$ limit. 
The parameter $\epsilon$ helps to organize the WKB series; 
we set $\epsilon =1$ when the calculation is completed. 
First we put
\beq
T(\theta )= \exp{ \big( {1\over \epsilon} S(\theta ) \big) } \; , 
\eeq
where $S(\theta )$ is a complex function that satisfies the differential 
equation
\beq
S{'}^2(\theta ) + \epsilon S{''}(\theta ) 
+ \epsilon \cot{(\theta )} S{'}(\theta ) = Q(\theta) \; .
\eeq
The WKB expansion for the function $S(\theta )$ is given by
\beq
S(\theta ) = \sum_{k=0}^{\infty} \epsilon^k S_k(\theta ) \; , 
\eeq
and by comparing like powers of $\epsilon$ we obtain a recursion 
formula ($n>0$)
\beq
S{'}_0^2 = Q , \;\;\;\;
\sum_{k=0}^n S{'}_k S{'}_{n-k} + S{''}_{n-1} 
+ \cot{(\theta )} S{'}_{n-1} =0 \; .
\eeq
Straightforward calculations give for the first few terms
\beq
S{'}_0= - Q^{1\over 2} \; ,
\eeq
\beq
S{'}_1 = -{1\over 4} Q{'} Q^{-1} - {1\over 2} \cot{(\theta )} \; ,
\eeq
\beqa
S{'}_2 & = & -{1\over 32} Q{'}^2Q^{-5/2} 
-{1\over 8} {d\over d\theta} ({ Q{'} Q^{-3/2}}) 
- {1\over 8} \cot^2{(\theta )} Q^{-1/2} \nonumber \\
& - & {1\over 4} ({d\over d\theta} \cot{(\theta )}) Q^{-1/2} \; .
\eeqa 
The exact quantization of the wave function is given by 
\beq
\oint dS = \sum_{k=0}^{\infty} \oint dS_k = 2 \pi i \; n_{\theta} \; ,
\eeq
where we have now set $\epsilon =1$. This integral is a complex contour 
integral which encircles the two turning points on the real axis. 
Obviously, it is derived from the requirement of the uniqueness of the 
complex wave function $T$ [4,6]. 
\par
The zero order term is given by
\beq
\oint dS_0 = 2 i \int d\theta \sqrt{\lambda^2 - W(\theta )} 
=2\pi i (\lambda - n_{\phi}) \; ,
\eeq
and the first term reads 
\beq
\oint dS_{1} = -{1\over 4} \ln{Q}|_{contour} = - \pi i \; .  
\eeq
Evaluating $\ln{Q}$ once around the contour gives $4\pi i$ because 
the contour encircles two simple zeros of $Q$. 
\par
All the other odd terms vanish when integrated along the closed 
contour because they are exact differentials [4]. 
So the quantization condition can be written as
\beq
\sum_{k=0}^{\infty} \oint dS_{2k} = 2 \pi i (n_{\theta} 
+{1\over 2})  \; ,
\eeq 
and thus  it is a sum over even--numbered terms only. 
The next non--zero term is given by
\beqa
\oint dS_2 & = & - i \big( 
{1\over 12} {\partial^2 \over \partial (\lambda^2)^2} 
\int d\theta {W{'}^2(\theta )\over \sqrt{\lambda^2 - W(\theta )}} 
+ {1\over 2} {\partial \over \partial (\lambda^2)} 
\int d\theta {W{'}(\theta) \cot{(\theta )} \over \sqrt{\lambda^2 - W(\theta )}} 
\nonumber \\
& + & {1\over 4} \int d\theta {\cot^2{(\theta )} \over 
\sqrt{\lambda^2 - W(\theta )}} \big) \; .
\eeqa
In all integrals the limits of integration are
between the two turning points. 
After substitution $z=\tan{(\theta )}$, we have
\beqa
\int d\theta {W{'}^2(\theta )\over \sqrt{\lambda^2 - W(\theta )}} 
& = & {4 n_{\phi}^4\over \sqrt{\lambda^2 - n_{\phi}^2} }
\int dz {(1+z^2)\over z^6} \sqrt{z^2\over z^2 - \beta } = \nonumber \\
& = & {3 \pi \over 2 n_{\phi}} (\lambda^2 - n_{\phi})^2 
+ 2\pi n_{\phi}(\lambda^2 - n_{\phi}^2) \; ,
\eeqa
where $\beta = n_{\phi}^2 /(\lambda^2 - n_{\phi}^2 )$, so that 
\beq
{\partial^2 \over \partial (\lambda^2)^2} 
\int d\theta {W{'}^2(\theta )\over \sqrt{\lambda^2 - W(\theta )}} 
= {3 \pi \over n_{\phi}} \; .
\eeq 
For the other integrals we use the same procedure.
\beq
\int d\theta {W{'}(\theta )\cot{(\theta )}\over \sqrt{\lambda^2 - W(\theta )}} 
= -{2 n_{\phi}^2 \over \sqrt{\lambda^2 - n_{\phi}^2} }
\int dz {1\over z^4} \sqrt{z^2\over z^2 - \beta } 
= - {\pi \over n_{\phi}}(\lambda^2 - n_{\phi}^2) \; ,
\eeq
from which we obtain
\beq
{\partial \over \partial (\lambda^2)} 
\int d\theta {W{'}(\theta )\cot{(\theta )} 
\over \sqrt{\lambda^2 - W(\theta )}} = - {\pi \over n_{\phi}} \; .
\eeq 
The last integral gives
\beq
\int d\theta {\cot^2{(\theta )}\over \sqrt{\lambda^2 - W(\theta )}} 
= {1\over \sqrt{\lambda^2 - n_{\phi}^2} }
\int dz {1\over z^2(1+z^2)} \sqrt{z^2\over z^2 - \beta } 
= \pi ({1\over n_{\phi}} - {1\over \lambda}) \; .
\eeq
In conclusion 
\beq
\oint dS_2 = - i \big( {1\over 12}{3 \pi \over n_{\phi}} + 
{1\over 2}(-{\pi \over n_{\phi}}) 
+ {1\over 4}\pi({1\over n_{\phi}} - {1\over \lambda}) \big) =
{\pi i \over 4 \lambda} \; ,
\eeq
where, importantly,  the $n_{\phi}$ dependence drops out now. 
Thus up to the second order in $\epsilon$ the quantization condition reads
\beq
\lambda + {1\over 8 \lambda } = l + {1\over 2} \; ,
\eeq
where $l=n_{\theta}+n_{\phi}$. The term $1/8\lambda$ is the first quantum 
correction to the the quantization of the angular momentum. 
From this result we can argue ("conjecture by educated guess") 
that the $\epsilon^{2k}$ term in the WKB series is ($k>0$) 
\beq
\oint dS_{2k} = 2\pi i { {1\over 2}\choose k} 2^{-2k} \lambda^{1-2k} \; ,
\eeq
so that the WKB expansion of the 
angular momentum to all orders is given by
\beq
\sum_{k=0}^{\infty} { {1\over 2}\choose k} 2^{-2k} \lambda^{1-2k} 
=l+{1\over 2} \; .
\eeq
This is the exact formula for the relationship between $l$ and
$\lambda$, because 
\beq
\sum_{k=0}^{\infty} { {1\over 2}\choose k} 2^{-2k} \lambda^{1-2k} 
={1\over 2}\sqrt{1 + 4 \lambda^2} \; ,
\eeq
and the equation $\sqrt{1 + 4 \lambda^2} /2 =l+1/2$ can be inverted and 
gives $\lambda =\sqrt{l(l+1)}$. This completes our investigation of the
semiclassical expansion for the angular momentum, where it remains
in general to prove the conjectured formula for $k \geq 2$.

\section{Discussion and conclusions}

In the present paper we offer the first
calculation of the higher WKB terms beyond the torus quantization 
leading terms for the angular momentum. 
\par
This analysis explains the exactness of the torus quantization 
for the entire 3-dim Kepler problem. As is well known, the 3-dim 
Kepler problem can be reduced to a 1-dim radial problem with 
potential $V(x)={L^2\over 2 x^2} -{1 \over x}$, where $L$ is 
the angular momentum. Since the problem is separable, 
the wave functions (for the angular momentum and for the radial part) 
multiply and their phases
have the additivity property, and therefore the total
phase written as $\frac{i}{\hbar} (\sigma - i\hbar S)$ must obey
the quantization condition (uniqueness of the wave function). 
As shown by the authors in [8], the quantum 
corrections (i.e. terms higher than the torus quantization terms)
do indeed compensate mutually term by term, and only the torus 
quantization terms remain.    
\par
One important future project is to analyze a more general class of the 
1-dim potentials and to extend results to integrable but not separable 
systems with two or more degrees of freedom. 

\section*{References} 

\parindent=0. pt

[1] A. Einstein, {\it Verh. Dtsch. Phys. Ges.} {\bf 19} 82 (1917)

[2] V.P. Maslov, {\it J Comp. Math. and Math. Phys.} {\bf 1} 113--128; 
638--663 (1961) (in Russian)

[3] V.P. Maslov and M.V. Fedoriuk, {\it Semi-Classical Approximations in 
Quantum Mechanics} (Boston: Reidel Publishing Company) (1981)

[4] C.M. Bender, K. Olaussen and P.S. Wang, 
{\it Phys. Rev. } D {\bf 16} 1740 (1977)

[5] E. Narimanov, private communication (1995)

[6] J.L. Dunham, {\it Phys. Rev.} {\bf 41} 713 (1993)

[7] A. Voros, {\it Ann. Inst. H. Poincar\`e} A {\bf 39} 211 (1983)

[8] M. Robnik and L. Salasnich, 
"WKB expansion for the angular momentum and the Kepler problem: from the 
torus quantization to the exact one", Preprint University of Maribor, 
CAMTP/96-4

\end{document}